\documentclass[3p,times,procedia]{elsarticle}
\usepackage{nupha_ecrc}

\volume{00}

\firstpage{1}

\journalname{Nuclear Physics A}

\runauth{}

\jid{nupha}

\jnltitlelogo{Nuclear Physics A}

\usepackage{amssymb}
\usepackage[figuresright]{rotating}

\begin{document}
\newcommand{\vAi}{{\cal A}_{i_1\cdots i_n}}
\newcommand{\vAim}{{\cal A}_{i_1\cdots i_{n-1}}}
\newcommand{\vAbi}{\bar{\cal A}^{i_1\cdots i_n}}
\newcommand{\vAbim}{\bar{\cal A}^{i_1\cdots i_{n-1}}}
\newcommand{\htS}{\hat{S}}
\newcommand{\htR}{\hat{R}}
\newcommand{\htB}{\hat{B}}
\newcommand{\htD}{\hat{D}}
\newcommand{\htV}{\hat{V}}
\newcommand{\cT}{{\cal T}}
\newcommand{\cM}{{\cal M}}
\newcommand{\cMs}{{\cal M}^*}
\newcommand{\vk}{\vec{\mathbf{k}}}
\newcommand{\bk}{\bm{k}}
\newcommand{\kt}{\bm{k}_\perp}
\newcommand{\kp}{k_\perp}
\newcommand{\km}{k_\mathrm{max}}
\newcommand{\vl}{\vec{\mathbf{l}}}
\newcommand{\bl}{\bm{l}}
\newcommand{\bK}{\bm{K}}
\newcommand{\bb}{\bm{b}}
\newcommand{\qm}{q_\mathrm{max}}
\newcommand{\vp}{\vec{\mathbf{p}}}
\newcommand{\bp}{\bm{p}}
\newcommand{\vq}{\vec{\mathbf{q}}}
\newcommand{\bq}{\bm{q}}
\newcommand{\qt}{\bm{q}_\perp}
\newcommand{\qp}{q_\perp}
\newcommand{\bQ}{\bm{Q}}
\newcommand{\vx}{\vec{\mathbf{x}}}
\newcommand{\bx}{\bm{x}}
\newcommand{\tr}{{{\rm Tr\,}}}
\newcommand{\bc}{\textcolor{blue}}

\newcommand{\beq}{\begin{equation}}
\newcommand{\eeq}[1]{\label{#1} \end{equation}}
\newcommand{\ee}{\end{equation}}
\newcommand{\bea}{\begin{eqnarray}}
\newcommand{\eea}{\end{eqnarray}}
\newcommand{\beqar}{\begin{eqnarray}}
\newcommand{\eeqar}[1]{\label{#1}\end{eqnarray}}

\newcommand{\half}{{\textstyle\frac{1}{2}}}
\newcommand{\ben}{\begin{enumerate}}
\newcommand{\een}{\end{enumerate}}
\newcommand{\bit}{\begin{itemize}}
\newcommand{\eit}{\end{itemize}}
\newcommand{\ec}{\end{center}}
\newcommand{\bra}[1]{\langle {#1}|}
\newcommand{\ket}[1]{|{#1}\rangle}
\newcommand{\norm}[2]{\langle{#1}|{#2}\rangle}
\newcommand{\brac}[3]{\langle{#1}|{#2}|{#3}\rangle}
\newcommand{\hilb}{{\cal H}}
\newcommand{\pleft}{\stackrel{\leftarrow}{\partial}}
\newcommand{\pright}{\stackrel{\rightarrow}{\partial}}
\newcommand{\dif}{\mathrm{d}}
\newcommand{\pT}{p_\perp}
\newcommand{\sNN}{\sqrt{s_{\mathrm{NN}}}}
\newcommand{\RAA}{R_{\mathrm{AA}}}
\begin{frontmatter}



%% Instructions from Editor: Please use the following \dochead only in the preprint version (e-print arXiv etc.); 
%% use empty \dochead{} when submitting to Nuclear Physics A!
\dochead{XXVIIIth International Conference on Ultrarelativistic Nucleus-Nucleus Collisions\\ (Quark Matter 2019)}
%% e.g. \dochead{17th International Conference on Dynamical Processes in Excited States of Solids}

\title{From high $p_\perp$ theory and data to inferring anisotropy of Quark-Gluon Plasma}


\author{Magdalena Djordjevic$^a$, Stefan Stojku$^a$, Dusan Zigic$^a$, Bojana Ilic$^a$, Jussi Auvinen$^a$, Igor Salom$^a$, Marko Djordjevic$^b$ and Pasi Huovinen$^a$}
\address{$^a$ Institute of Physics Belgrade, University of Belgrade, Serbia}
\address{$^b$ Faculty of Biology, University of Belgrade, Serbia}

\begin{abstract}
High $p_\perp$ theory and data are commonly used to study  high $p_\perp$ parton interactions with QGP, while low $p_\perp$ data and corresponding models are employed to infer QGP bulk properties.  
On the other hand, with a proper description of high  $p_\perp$ parton-medium interactions, high $p_\perp$ probes become also powerful tomography tools, since they are sensitive to global QGP features, such as different temperature profiles or initial conditions. This tomographic role of high $p_\perp$ probes can be utilized to assess the spatial anisotropy of the QCD matter. 
With our dynamical energy loss formalism, we show
that a (modified) ratio of $R_{AA}$ and $v_2$ presents a reliable and robust observable for straightforward extraction of initial state anisotropy. We analytically estimated the proportionality between the $v_2/(1-R_{AA})$ and anisotropy coefficient $\epsilon_{2L}$, and found surprisingly good agreement with full-fledged numerical calculations. Within the current error bars, the  extraction of the anisotropy from the existing data using this approach is still inaccessible. However, with the expected accuracy improvement in the upcoming LHC runs, the anisotropy of the QGP formed in heavy ion collisions can be straightforwardly derived  from the data. Such a data-based anisotropy parameter would present an important test to models describing the initial stages of heavy-ion collision and formation of  QGP, and demonstrate the usefulness of high $p_\perp$ theory and data in obtaining QGP properties.
\end{abstract}


\end{frontmatter}

% \linenumbers

\section{Introduction}
\label{}
Understanding the properties of the new form of matter named  Quark-Gluon Plasma (QGP) is the major goal of relativistic heavy ion physics~\cite{QGP_1,QGP_2}. However, to explore the properties of QGP, one needs good probes. With regards to that, it is commonly assumed that high $p_\perp$  theory and data are good probes for exploring the high $p_\perp$  parton interactions with QGP, while low $p_\perp$ theory and data are considered as good probes for bulk QGP properties. Contrary to this common assumption, the goal of this contribution is to demonstrate that high $p_\perp$ particles can also be useful independent probes of {\it bulk} QGP properties. 

To put it simply, the main idea is that when high $p_\perp$ particles transverse QGP, they lose energy, where this energy loss is sensitive to bulk QGP properties, such as its temperature profiles or initial conditions. Consequently, with a realistic and sophisticated high $p_\perp$ parton energy loss model, high $p_\perp$ probes can indeed become powerful tomographic tools. So, in this contribution, we will present how we can use these probes to infer some of the bulk QGP properties, i.e., for precision QGP tomography. Note that only the main results are presented here; for a more detailed version, see~\cite{Asymmetry}, and references therein.

\section{DREENA framework}
To achieve the goal of utilizing high $p_\perp$ theory and data for inferring the bulk QGP properties, as previously implied, a reliable high $p_\perp$ parton energy loss model is necessary. With this goal in mind, we developed a dynamical energy loss formalism~\cite{MD_Dyn,MD_Col}, which takes into account some more realistic and unique features, such as: i) The calculations are performed within finite  temperature field theory and generalized Hard-Thermal-Loop~\cite{Kapusta} approach, in which the infrared divergences are naturally regulated, excluding the need for artificial cutoffs. ii) The formalism assumes QCD medium of finite size and finite temperature, consisting of dynamical partons (i.e., energy exchange with medium constituents is included), in distinction  to commonly considered static scatterers approximation and/or models with vacuum-like propagators.  iii) Both radiative~\cite{MD_Dyn} and collisional~\cite{MD_Col} energy losses are calculated within the same theoretical framework, and are equally applicable to light and heavy flavors. iv) The formalism is generalized to include a finite chromomagnetic mass~\cite{MD_MagnMass}, running coupling, and to relax the widely used soft-gluon approximation~\cite{BDD}. Finally, the formalism is integrated in a numerical framework DREENA (Dynamical Radiative and Elastic ENergy loss Approach)~\cite{DREENA_C,DREENA_B}, to provide predictions for high $p_\perp$ observables.

Within this framework, we generated a wide set of high $p_\perp$ predictions using 1D Bjorken expansion~\cite{BjorkenT}  (i.e., DREENA-B framework~\cite{DREENA_B}). Thus we obtained a good joint agreement with a wide range of high $p_\perp$ $R_{AA}$  and $v_2$ data, by applying the same numerical procedure, the same parameter set, and no fitting parameters in model testing. That is, there is no $v_2$ puzzle~\cite{v2_puzzle} within our model, which then strongly suggests that the model provides a realistic description of high $p_\perp$ parton-medium interactions. Moreover, our preliminary findings suggest that, within our formalism, moving from 1D Bjorken to full 3D hydrodynamical expansion does not significantly affect the agreement of our predictions with high $p_\perp$ $R_{AA}$ and $v_2$ data~\cite{ZigicPoster}. Consequently, in order to adequately address the high $p_\perp$ measurements, a proper description of high $p_\perp$ parton interactions with the medium appears to be much more important than an advanced medium evolution description. 
Furthermore, we have also analyzed the sensitivity of high  $p_\perp$ $R_{AA}$ and $v_2$ to different initial stages, %where we obtained that high  $p_\perp$ $R_{AA}$ shows notable sensitivity to initial conditions, while $v_2$ predictions are surprisingly inensitive to different initial stages, 
giving an additional insigth in the usefulness of both high $p_\perp$ observables in the precision QGP tomography~\cite{ISPaper}.

\section{Inferring QGP anisotropy through high $p_\perp$ theory and data}

As one example of QGP tomography, in this contribution, we will address how to infer the QGP anisotropy from high $p_\perp$ $R_{AA}$ and $v_2$ data. The initial state anisotropy is one of the main properties of QGP and a major limiting factor for precision QGP tomography.
However, despite its essential importance, it is still not possible to directly infer the initial anisotropy from experimental measurements. 
Several theoretical studies~\cite{MCGlauber,EKRT,IPGlasma,MCKLN} have provided different methods for calculating the initial anisotropy, leading to notably different predictions, with a notable effect in the resulting predictions for both low and high  $p_\perp$ data. 
Therefore, approaches for inferring anisotropy from the data are necessary. Optimally, these approaches should be complementary to existing predictions, i.e., based on a method that is fundamentally different from models of early stages of QCD matter.

To this end, we here propose a novel approach to extract the initial state anisotropy. Our method is based on inference from high  $p_\perp$ data, by using already available $R_{AA}$ and $v_2$ measurements, which will moreover be measured with much higher precision in the future. Such an approach is substantially different from the existing approaches, as it is based on the inference from experimental data (rather than on calculations of early stages of QCD matter) exploiting the information from interactions of rare high $p_\perp$ partons with the QCD medium. This also presents an improvement/optimization in utilizing high $p_\perp$ data as, to date, these data were mostly constrained on studying the parton-medium interactions, rather than assessing bulk QGP parameters, such as spatial asymmetry.

In the literature, the initial state anisotropy is quantified in terms of eccentricity parameter $\epsilon_2$
\begin{equation}
 \epsilon_2 = \frac{\langle y^2-x^2 \rangle}{\langle y^2+x^2 \rangle}
            = \frac{\int\dif x\,\dif y\, (y^2-x^2)\, \rho(x,y)}
                   {\int\dif x\,\dif y\, (y^2+x^2)\, \rho(x,y)},
 \label{eccentricity}
\end{equation}
where $\rho(x,y)$ denotes the initial density distribution of the formed QGP. Regarding high  $p_\perp$ observables, we note that $v_2$ is sensitive to both the anisotropy of the system and its size, while $R_{AA}$ is sensitive only to the size of the system. Therefore, it is plausible that the adequate  observable for extracting eccentricity from high  $p_\perp$ data depends on both $v_2$ and $R_{AA}$, and the question is how.

To address this question, we will use the dynamical energy loss formalism, and DREENA-B framework outlined above. For high  $p_\perp$, the fractional energy loss scales as~\cite{Asymmetry}
$ \Delta E/E \sim \chi \langle T \rangle^a \langle L \rangle^b$,
where $\langle T \rangle$ stands for the average temperature along the path of high $p_\perp$ parton, $\langle L \rangle$ is the average path-length traversed by the
parton, $\chi$ is a proportionality factor that depends on the initial parton transverse momentum, and $a$ and $b$ are exponents which
govern the temperature and path-length dependence of the energy
loss. Within our model, $a \approx 1.2$ and $b \approx 1.4$, which is contrary to simpler models, and consistent with a wide range of experimental data~\cite{MD_5TeV,NewObservable}.
From this simple scaling argument, we can straightforwardly obtain the following expressions for $R_{AA}$ and $v_2$ (for more details we refer the reader to~\citep{Asymmetry}):
\begin{eqnarray}
 \RAA \approx 1-\xi(\chi) \langle T \rangle^a \langle L \rangle^b, \;\  \ \ \ \ \ \ \ \ \,\,\,\,\,\,\, v_{2} \approx  \frac{1}{2} \frac{\RAA^{in} - \RAA^{out}}{\RAA^{in} + \RAA^{out}}
          \approx  \xi(\chi) \langle T\rangle^a \langle L\rangle^b
                     \left( \frac{b}{2} \frac{\Delta L}{\langle L \rangle}
                           - \frac{a}{2}\frac{\Delta T}{\langle T \rangle} \right),
\label{v2approx}
\end{eqnarray}
 where we see that $\xi(\chi) \langle T \rangle^a \langle L \rangle^b$ corresponds to $1-R_{AA}$. Therefore, if we divide $v_2$ by ($1-R_{AA}$), we see that this ratio is given by the following simple expression:
\begin{eqnarray}
  \frac{v_{2}}{1-\RAA} \approx \left( \frac{b}{2} \frac{\Delta L}{\langle L \rangle}
                                   - \frac{a}{2}\frac{\Delta T}{\langle T \rangle} \right).
\label{v2RaaRatio}
\end{eqnarray}
Note that, while this ratio exposes the dependence  on the asymmetry of the system (through spatial $(\Delta L / \langle L \rangle)$ and temperature $(\Delta T /\langle T \rangle)$ parts), the dependence only on  spatial anisotropy is still not isolated. %(singled out, isolated). 
 However, by plotting together spatial and temperature anisotropy, we obtain a linear dependence~\cite{Asymmetry}, with a proportionality factor given by $c\approx4.3$. Therefore, $v_2/(1-R_{AA})$ reduces to the following expression:
\begin{eqnarray}
  \frac{v_{2}}{1-\RAA} \approx\frac{1}{2} \left(b - \frac{a}{c}\right)
       \frac{\langle L_{out}\rangle - \langle L_{in} \rangle}
            {\langle L_{out}\rangle + \langle L_{in} \rangle}
   \approx 0.57 \varsigma ,   \ \ \ \ \                                
{\rm where} \, \; \;  \varsigma = \frac{\langle L_{out}\rangle - \langle L_{in} \rangle}
                                           {\langle L_{out}\rangle + \langle L_{in} \rangle}
\; \; \; {\rm and} \;\; \; \frac{1}{2} (b - \frac{a}{c}) \approx 0.57. \; \; \; \; \; \;
\label{AsymetryEq0}
\end{eqnarray}
Consequently, the asymptotic scaling behavior of observables $v_2$ and $R_{AA}$, at high  $p_\perp$, reveals that their (moderated) ratio is determined only by the geometry of the initial QGP droplet. Therefore, the anisotropy parameter $\varsigma$ could, in principle, be directly obtained from the high $p_\perp$ experimental data.

\begin{figure}[h]
\includegraphics[scale=0.45]{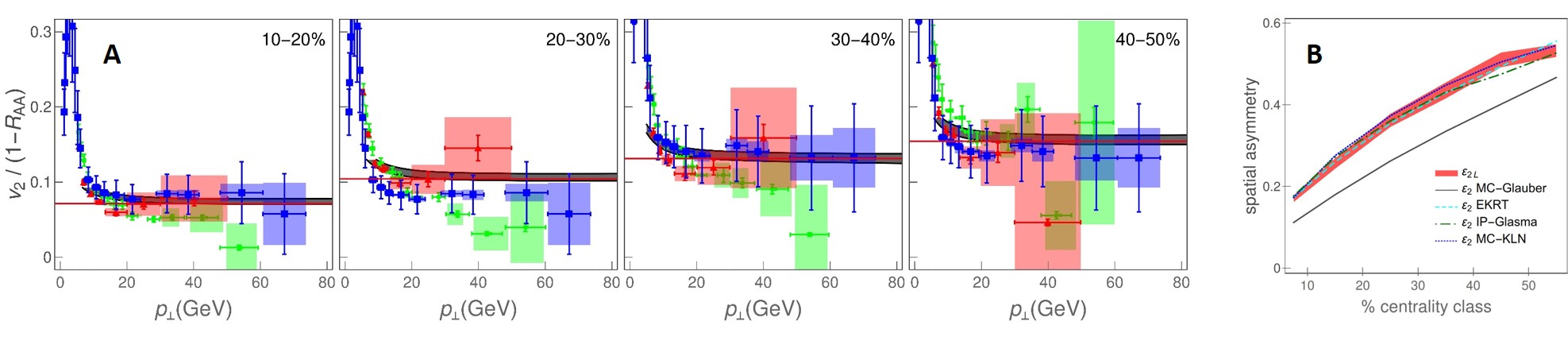}
\vspace*{-0.8cm}
\caption{{\bf A)} Comparison of theoretical predictions for charged hadron $v_{2}/(1-R_{AA})$ as a function of $p_{\perp}$  with $5.02$ TeV
  $Pb+Pb$  CMS~\cite{CMS_CH_RAA,CMS_CH_v2} (blue squares), ALICE~\cite{ALICE_CH_RAA,ALICE_CH_v2} (red triangles) and ATLAS~\cite{ATLAS_CH_RAA,ATLAS_CH_v2} (green circles) data. Each panel corresponds to different centrality range, as indicated in the upper right corners, while red lines denote the limit $0.57 \varsigma$ from Eq.~(\ref{AsymetryEq0}). {\bf B)} Comparison of $\epsilon_{2 L}$ (red band) extracted from our full-fledged calculations, with $\epsilon_2$ obtained from MC-Glauber~\cite{MCGlauber} (gray full curve), EKRT~\cite{EKRT} (cyan dashed curve), IP-Glasma~\cite{IPGlasma} (green dot-dashed curve) and MC-KLN~\cite{MCKLN} (blue dotted curve) models. MC-Glauber and EKRT curves correspond to 5.02 TeV, whereas IP-Glasma and MC-KLN curves correspond to 2.76 TeV $Pb+Pb$ collisions at the LHC.}
\label{ExpComparisonFig}
\end{figure}
To test the adequacy of the analytical estimate given by Eqs.~(\ref{v2approx})-(\ref{AsymetryEq0}), Fig.~\ref{ExpComparisonFig}A is displayed, which comprises our $v_2/(1-R_{AA})$ predictions (gray bands), stemming from our full-fledged recently developed DREENA-B framework (outlined in the previous section), the ALICE, CMS and ATLAS data, and analytically derived asymptote $0.57 \varsigma$ (red lines). 
 Importantly, for each centrality range and for  $p_\perp \gtrsim 20$ GeV, $v_2/(1-R_{AA})$ is independent on $p_\perp$, and approaches the asymptote, i.e., is determined by the geometry of the system - depicted by the solid red line, up to $5\%$ accuracy.  Moreover, the experimental data for all three experiments also display the independence on the $p_\perp$ and agree  with our predictions, although the error bars are rather large. Therefore, we conclude that our scaling estimates are valid and that $v_2/(1-R_{AA})$ indeed carries the information about the anisotropy of the fireball, which can be simply (from the straight line fit to data at high $p_\perp$ limit) and robustly (in the same way for each centrality) inferred from the experimental data.

However, note that the anisotropy parameter $\varsigma$ is not the widely-considered anisotropy parameter $\epsilon_2$ (given by Eq.~(\ref{eccentricity})). To facilitate comparison with $\epsilon_2$  values in the literature, we define $\epsilon_{2L} = \frac{\langle L_{out}\rangle^2 - \langle L_{in}\rangle^2}
                  {\langle L_{out}\rangle^2 + \langle L_{in}\rangle^2}
             = \frac{2 \varsigma}{1+\varsigma^2}$, and in Fig.~\ref{ExpComparisonFig}B compare it with the results from different initial-state models~\citep{MCGlauber,EKRT,IPGlasma,MCKLN}. First, we should note that as a starting point, our initial $\epsilon_2$, through which we generate our path-length distributions, agrees with EKRT and IP-Glasma. However, what is highly non-trivial is that, as an outcome of this procedure, in which $v_2/(1-R_{AA})$ is calculated (based on the full-fledged DREENA-B framework), we obtain $\epsilon_{2L}$ which practically  coincides with our initial $\epsilon_2$ and also with some of the conventional initial-state models.  %Note that $\epsilon_2$ is indirectly introduced in $R_{AA}$ and $v_2$ calculations through path-length distributions, while our calculations are performed using full-fledged numerical procedure.%
    As an overall conclusion, the straightforward extraction of $\epsilon_{2L}$ and its agreement with values of the prevailing initial-state models' eccentricity (and our initial $\epsilon_2$) is highly non-trivial and supports $v_2/(1-R_{AA})$ as a reliable and robust observable for anisotropy. Additionally, the width of our $\epsilon_{2L}$ band is smaller than the difference in the $\epsilon_2$ values obtained by using different models (e.g., MC-Glauber vs. MC-KLN). Therefore, our approach provides genuine resolving power to distinguish between different initial-state models, although it may not be possible to separate the finer details of more sophisticated models. This resolving power, moreover, comes from an entirely different perspective, i.e., from high  $p_\perp$ theory and data, supporting the usefulness of utilizing high $p_\perp$ theory and data for inferring the bulk QGP properties.

\bigskip
{\bf Acknowledgements:}
This work is supported by the European Research Council, grant ERC-2016-COG: 725741, and by the Ministry of Science and Technological
Development of the Republic of Serbia, under project numbers ON171004, ON173052 and ON171031.

\end{document}